\documentclass{elsart}

\usepackage{natbib}

\usepackage{graphicx}
\usepackage{amssymb}

\begin{document}

\newcommand{\HI} {\rm H\,{\rm I}}
\newcommand{\HII} {\rm H\,{\rm II}}
\newcommand{\Ha} {$\rm H \alpha$}
\newcommand{\kms} {\,${\rm km\,s}^{-1}$}
\newcommand{\kmskpc} {\,${\rm km\,s^{-1}\,kpc^{-1}}$}
\newcommand{\kpc} {$\,{\rm kpc}$}
\newcommand{\de}{$^{\circ}$}
\newcommand{\Mo}{$\,{M}_\odot$}
\newcommand{\yr}{{\rm yr}}
\newcommand{\moyr}{${M_\odot\,\rm yr}^{-1}$}
\newcommand{\mopc}{$M_\odot\,{\rm pc^{-2}}$}
\newcommand{\lo}{L_{\odot}}
\newcommand{\loB}{L_{\odot \rm, B}}
\newcommand{\gsim}{\lower.7ex\hbox{$\;\stackrel{\textstyle>}{\sim}\;$}}
\newcommand{\lsim}{\lower.7ex\hbox{$\;\stackrel{\textstyle<}{\sim}\;$}}
\newcommand{\ergs}{\,{\rm erg\,s}^{-1}}

\begin{frontmatter}

\title{Gaseous Haloes: Linking Galaxies to the IGM}

\author[Bo1,Bo2]{Filippo Fraternali}
\ead{filippo.fraternali@unibo.it}
\author[Ox]{James Binney}
\author[Dw,Ka]{Tom Oosterloo}
\author[Bo2,Ka]{Renzo Sancisi}
\address[Bo1]{Department of Astronomy, University of Bologna, via Ranzani 1, 40127, Bologna, Italy}
\address[Ox]{Rudolf Peierls Centre for Theoretical Physics, 1 Keble Rd, Oxford, OX1 3NP, UK}
\address[Dw]{ASTRON, Postbus 2, 7990 AA Dwingeloo, NL}
\address[Bo2]{INAF - Astronomical Observatory, via Ranzani 1, 40127, Bologna, Italy}
\address[Ka]{Kapteyn Astronomical Institute, Postbus 800, 9700 AV Groningen, NL}

\begin{abstract}
In recent years evidence has accumulated that nearby spiral galaxies 
are surrounded by massive haloes of neutral and ionised gas.
These gaseous haloes rotate more slowly than the disks and show 
inflow motions. 
They are clearly analogous to the High Velocity Clouds of the Milky Way.
We show that these haloes cannot be produced by a galactic fountain process
(supernova outflows from the disk) where the fountain gas conserves
its angular momentum.
Making this gas interact with a pre-existing hot corona 
does not solve the problem.
These results point at the need for a substantial accretion of low 
angular momentum material from the IGM.
\end{abstract}


\end{frontmatter}

% main text
\section{Introduction}
\label{intro}

Recent deep neutral hydrogen (\HI) observations of several
nearby spiral galaxies (e.g.\ NGC\,891, \citet{swa97};
NGC\,2403, \citet{fra02};
M31, \citet{wes05}; UGC\,7321, \citet{mat03})
indicate that a large fraction of the neutral gas is located outside
the plane of the disk.
Such ``halo'' gas has peculiar kinematics: it rotates more slowly
than the gas in the disk \citep{swa97} and shows
strong vertical motions from and/or towards the disk
\citep{boo05}.
In two galaxies (NGC\,2403, \citet{fra01}; and NGC\,4559, \citet{barb05})
the halo gas also shows an
overall radial inflow towards the centre of the galaxy.
Halo gas is also detected in optical emission lines
\citep[e.g.][]{hoo99, ros03} with kinematics similar to that of the
neutral gas \citep{hea06b} 
and in X-rays \citep[e.g.][]{str04}.
Halo gas around external galaxies has the same properties (masses and
velocities) of 
the Intermediate and High Velocity Clouds (IVCs and HVCs) of the Milky
Way \citep{wak97}.

What is the origin of these extended gaseous haloes?
In galaxies like NGC\,891 gas
is ejected from the disk by stellar activity producing a circulation
called galactic fountain \citep{sha76, bre80}.
However, it is now clear that a ``pure'' (ballistic) galactic fountain
cannot explain the kinematics of the halo gas \citep{fra06, hea06b}.
In particular there is a need for 
the fountain gas to loose part of its angular momentum.
Such a loss can be achieved in two ways: 1) interaction between the fountain
clouds and a pre-existing hot corona or 2) interaction with accreting material.
We explore here the first possibility.

\section{The Structure of the HI halo of NGC891}

\begin{figure}[ht]
\begin{center}
\includegraphics[width=300pt]{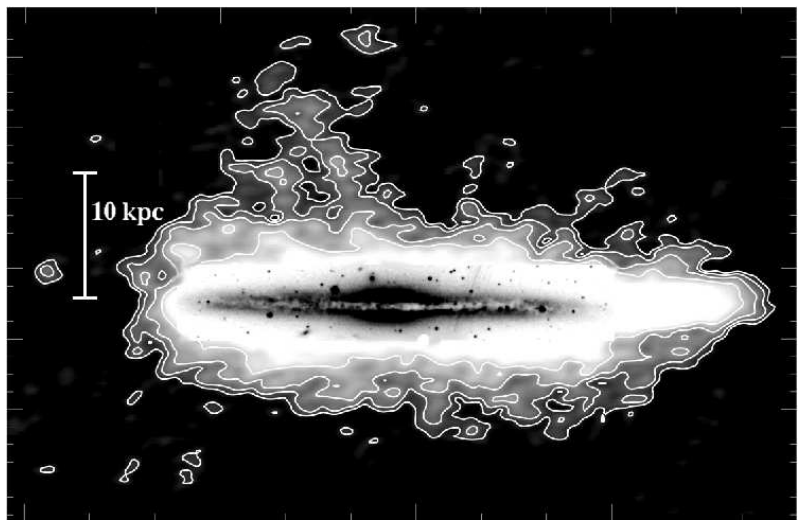}
\caption{
Optical DSS image (grey-scale) and total $\HI$ map 
(contours$+$negative grey-scale) of NGC\,891.
$\HI$ contours are: 1, 2, 4, 8, 16 $\times$
10$^{19}$ atoms cm$^{-2}$. The beam size is $25'' = $1.15 \kpc.
\label{f_totalHI}}
\end{center}
\end{figure}

New deep \HI\ observations of the nearby edge-on spiral galaxy NGC 891 
obtained with the Westerbork Synthesis Radio Telescope show an 
extended and massive (25\% of the total \HI\ mass) 
halo component of neutral gas \citep{oos07}.

Fig.\ \ref{f_totalHI} shows the total \HI\ map of NGC\,891 (countours
$+$ white shade) with 
overlaid an optical image (grey-scale).
The map is rotated 67 degrees counter-clock wise 
with respect to the on-sky orientation.
The neutral gas everywhere extends to large heights ($z\sim8\,$kpc) from the
plane.
Some features, in particular a long filament, extend to much larger 
distances ($\sim$20 kpc from the plane).
The kinematics of the halo gas in NGC\,891 is characterized by a regular
decrease in rotational velocity with $z$ 
(gradient of $\sim$15 \kmskpc) \citep{fra05}.
Most of the high latitude features, i.e.\ the filament, are at 
roughly the systemic velocity.

\section{Modelling the halo gas}

\citet{fra06} have presented a model for the formation of the 
halo gas in spiral galaxies.
The model is built by integrating the orbits of gas clouds that are ejected 
from the disk of a spiral galaxy, and then move ballistically up into the
halo and back down to the disk.
At each timestep,
the positions and velocities of the particles are projected
along the line of sight to produce an artificial cube, 
that can be compared with observed \HI\ data cubes.

\begin{figure}[ht]
\begin{center}
\includegraphics[width=300pt]{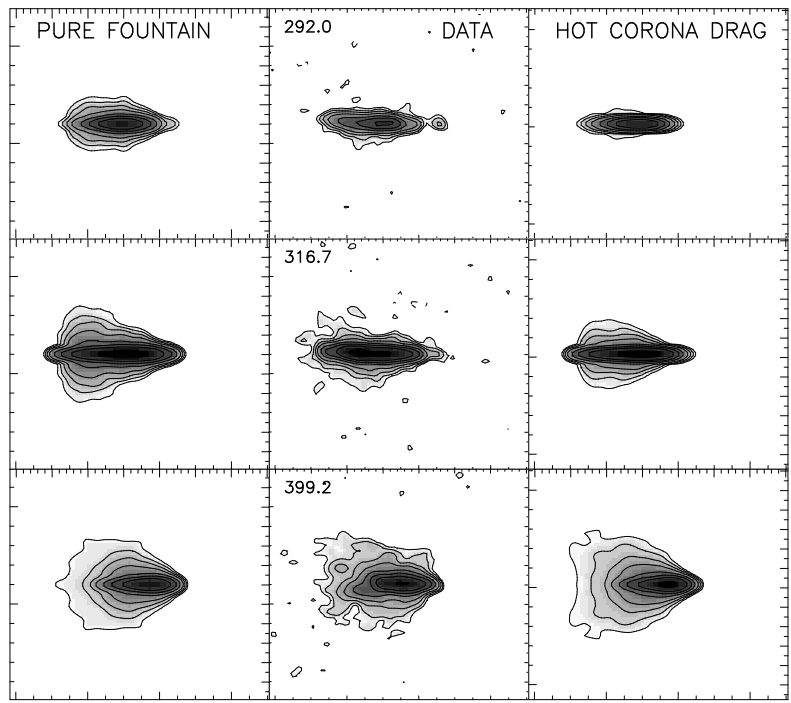}
\caption{Comparison between three representative channel maps for NGC\,891 (central column) and two models: pure galactic fountain (left column) and
galactic fountain interacting with the hot corona (right column).
Heliocentric velocities (km/s) are on the top left corners of the data.
\label{f_models}}
\end{center}
\end{figure}

\subsection{Pure galactic fountain}

At first we consider a model with only gravitational effects
(ballistic galactic fountain) \citep{fra06}.
Fig.\ \ref{f_models} shows three channel maps of NGC\,891 (central column)
compared with the prediction of such a fountain model (left column).
The top row shows a channel map at velocity far away from the 
systemic velocity ($\rm v_{sys}=$ 530 \kms).
Here the gas is rotating at high speed and it is all 
concentrated in the disk.
The more we move towards the systemic velocity the more the maps become
thicker;
in other words the halo gas rotates much more slowly than the disk gas.
Clearly the maps produced by the pure fountain model show more
fast-rotating halo gas than the data.

\subsection{Interaction with the hot corona}

We consider the interaction (drag) between the fountain gas and a pre-existing 
hot corona rotating more slowly than the disk (Fraternali \& Binney, in preparation).
We model the corona as a plasma at $T=2.7 \times 10^6\,{\rm K}$ 
in equilibrium in the axisymmetric potential.
The plasma density is normalized using the observed 
X-ray luminosity of NGC\,891 \citep{str04}.
Fig.\ \ref{f_models} (third column) shows the result of such a model with
a hot corona rotating a constant speed v$_{\phi}(R)=$ 100 \kms.
Clearly this produces an improvement with respect to the pure fountain model.

\begin{figure}[ht]
\begin{center}
\includegraphics[width=290pt]{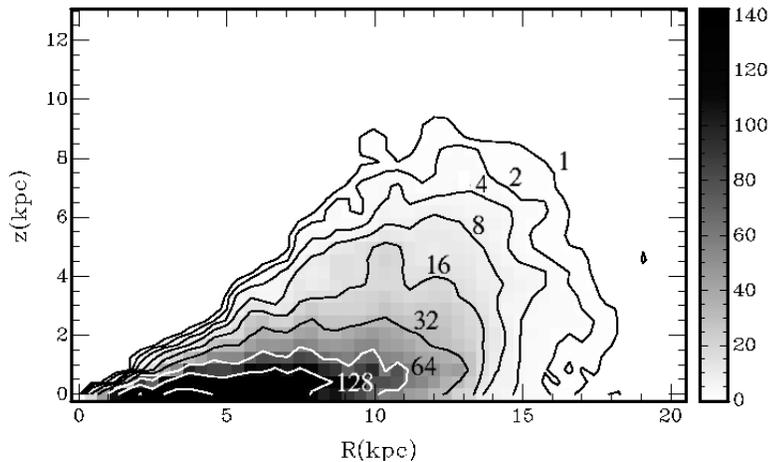}
\caption{Azimuthal velocities acquired by the hot corona per Myr
due to the transferring of angular momentum from the fountain gas.
\label{f_spinup}}
\end{center}
\end{figure}

The key point of the above model is that
the fountain particles transfer part of their angular momentum
to the hot corona.
However, can the corona absorb such momentum?
Fig.\ \ref{f_spinup} shows the azimuthal velocity acquired by the inner 
corona after a Myr as a function of the coordinates $R$ and $z$.
The spin-up time of the inner halo is very rapid; within a time t$_{\rm spinup}
\lsim$ 1 Myr the inner corona co-rotates with the disk and the drag vanishes.
Therefore the hot corona cannot be responsible for the observed 
lag of the halo gas.

\section{Conclusions}

We have seen that: 
1) the spiral galaxy NGC\,891
is surrounded by a massive halo of neutral gas
extending up to more than 8 kpc from the plane;
2) extraplanar gas such as that in NGC 891 seems to be common
in spiral galaxies and is analogous to the HVCs of the Milky Way;
3) models purely based on a galactic fountain, 
with or without interaction 
with the hot corona, fail to reproduce the data.
More promising models involve the interaction of the fountain
clouds with material accreted from the IGM (Fraternali \& Binney, in preparation).

\end{document}